# An inquiry-based laboratory on friction

Vera Montalbano

Department of Physical Sciences, Earth and Environment, University of Siena, Siena, Italy

Abstract – date of last change of the paper: 17.04.2014

*Sliding friction is usually introduced in high school, but rarely through activities in laboratory. A qualitative introduction to friction is presented by proposing exploration of different kinds of materials in order to suggest which aspects can be relevant and which interaction is involved. Different quantitative experiments are proposed for studying Leonardo's laws for friction. The learning path was tested with two high school classes during an instruction trip at department. Students were engaged in the inquiry-based introductory activity and seemed to realize with care the measurements. However, the analysis of their reports shows some learning difficulties.*

## Introduction

Friction is a complex phenomenon, usually missing in high school laboratory. Nevertheless, sliding friction is often introduced in the easier way in mechanics because of simple exercises which can be proposed on this subject. Another important aspect rarely explored in laboratory is the relation between static and kinetic friction.

Furthermore, a closer look in this topic allows to introduce interesting discussions on the structure of matter, the relations between macroscopic and microscopic modelling, which aspects of a phenomenon are really relevant, which others can be omitted and for which reason, and so on. Last but not least, friction can be the starting point for introducing students to the nanoscience world.

Relying on a recent study in higher education [1, 2], a learning path on friction had been designed for high school students. The first activity was a qualitative explorative path on friction designed for stimulating discussions in a workshop in which physics teachers and students were involved. The unexpected reactions, especially from teachers, highlighted the necessity to realize a wider research [3] with the aim of improving the learning on this topic both in secondary school and in higher education.

The activity had been realized within the Italian National Plan for Science Degrees [4-5].

### National Plan for Science Degrees

The decline of interest in studying science is a serious concern. In recent decades, a consistent decrease of graduates in science disciplines has been detected almost everywhere. In order to reverse this trend in Italy, the Ministry of Education and Scientific Research promoted since 2005 a wide national project (Progetto nazionale per le Lauree Scientifiche, i.e. PLS) [4].

The main actions of PLS were professional development for teachers and orientation for students, essentially by means of laboratory activities. In 2009, the National Plan for Science Degree (same acronym PLS) was launched and some of the most effective methodological aspects of the previous project were emphasized in new guidelines [4, 5] focused on:

- orienting to science degree by means of training
- the laboratory as a method not as a place
- the student must become the main character of learning
- favouring joint planning by teachers and university.

In this context, we offered the opportunity for secondary classes of performing some learning paths in educational laboratories at university.

**An introduction to friction in laboratory**

The learning path on friction started with an initial inquiry-based explorative laboratory. Common materials, such as wood, were examined and their behaviour was easily predicted. Another common material with a totally different behaviour was examined in order to induce a cognitive conflict. A final activity suggested the correct interaction involved in all kinds of friction.

A quantitative laboratory followed the qualitative exploration. The same materials could be used for verifying the Leonardo's laws for sliding friction.

The laboratory on friction was tested by two 3rd classes of a scientific high school (44 students, age 16 – 17 y), during an educational trip at physics department. Sliding friction and Leonardo's laws were introduced, discussed and assessed in previous school time by their physics teacher. In the next section, the activities designed for the laboratory are presented together with some remarks on the learning process. A discussion on results, some learning difficulties and suggestions for further activities are given in the last section.

## A qualitative introduction to friction

Students were invited to predict the behaviour of different sliding surfaces by using their previous knowledge and experience. Afterwards, they could realize and observe what really happened. New previsions were made and checked. Some hints were given by proposing an activity designed for selecting relevant aspects and the involved interaction. The students' description of the phenomenon changed during the qualitative path. All activities were realized with low-cost materials.

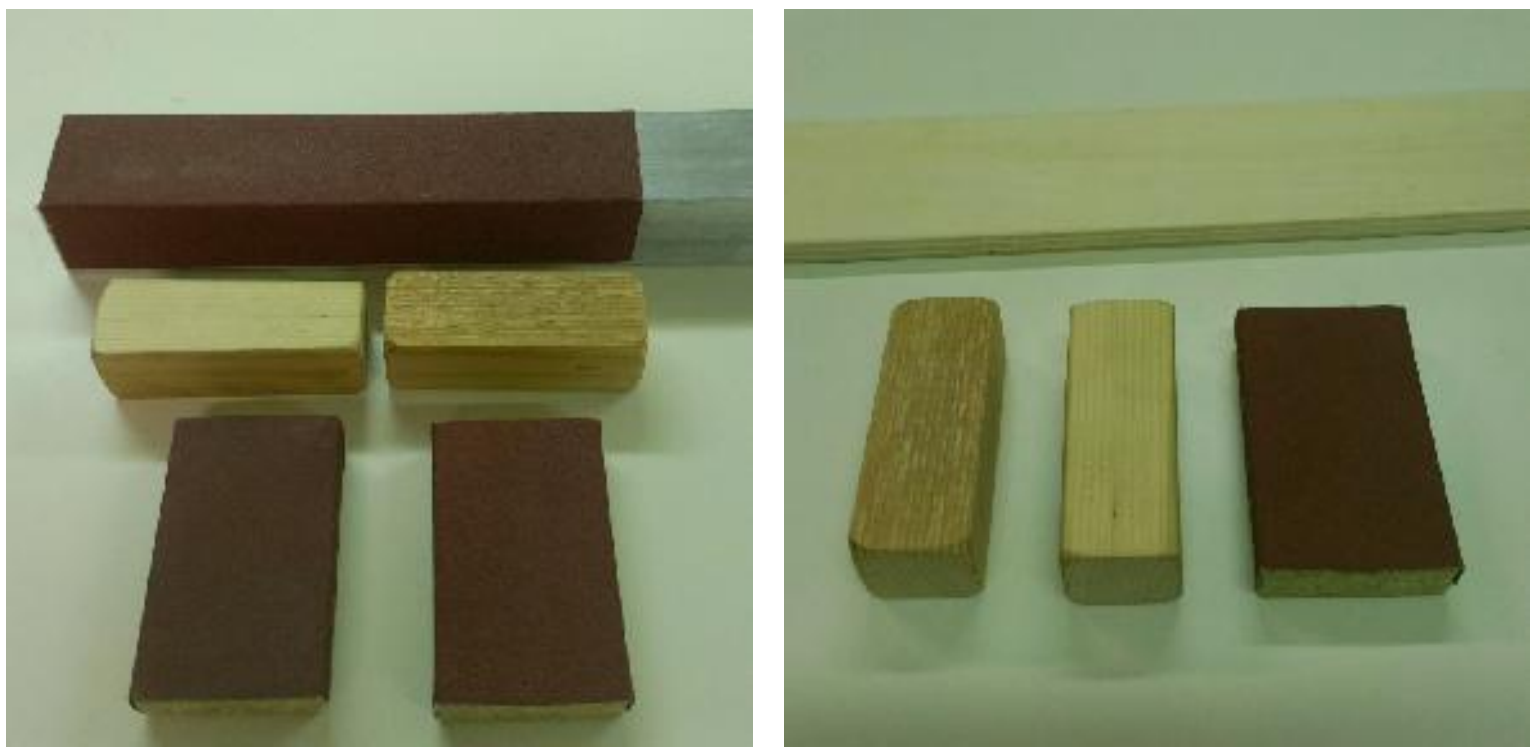

Figure 1. Different smooth and rough surfaces were explored.

Students were guided by a detailed worksheet and the qualitative exploration lasted about 90 minutes, during which teachers were available to provide clarification, students were involved in animated discussions and in seeking consistent answers to the questions that were gradually raising.

Smooth and rough wooden surfaces and different types of sandpaper, like those showed in fig. 1, were examined. Sandpaper and wood were touched, predictions on sliding behaviour for surfaces with the same and different material or roughness were requested and tested soon after. A qualitative graph on how the friction force intensity varies with the roughness of the sliding surfaces had been guessed and drawn (see fig. 2).

Surface roughness is a measure of the texture of a surface. It can be quantified by the vertical deviations of a real surface from its ideal plan. Students could made roughly comparative evaluations of this texture parameter based upon tactile sense intensity. Human tactile perception can discriminate to the nanoscale, as shown in recent research [6], and questions on this part of the exploration can be the starting point for following discussions on the relevance of the nanoscale in everyday life.

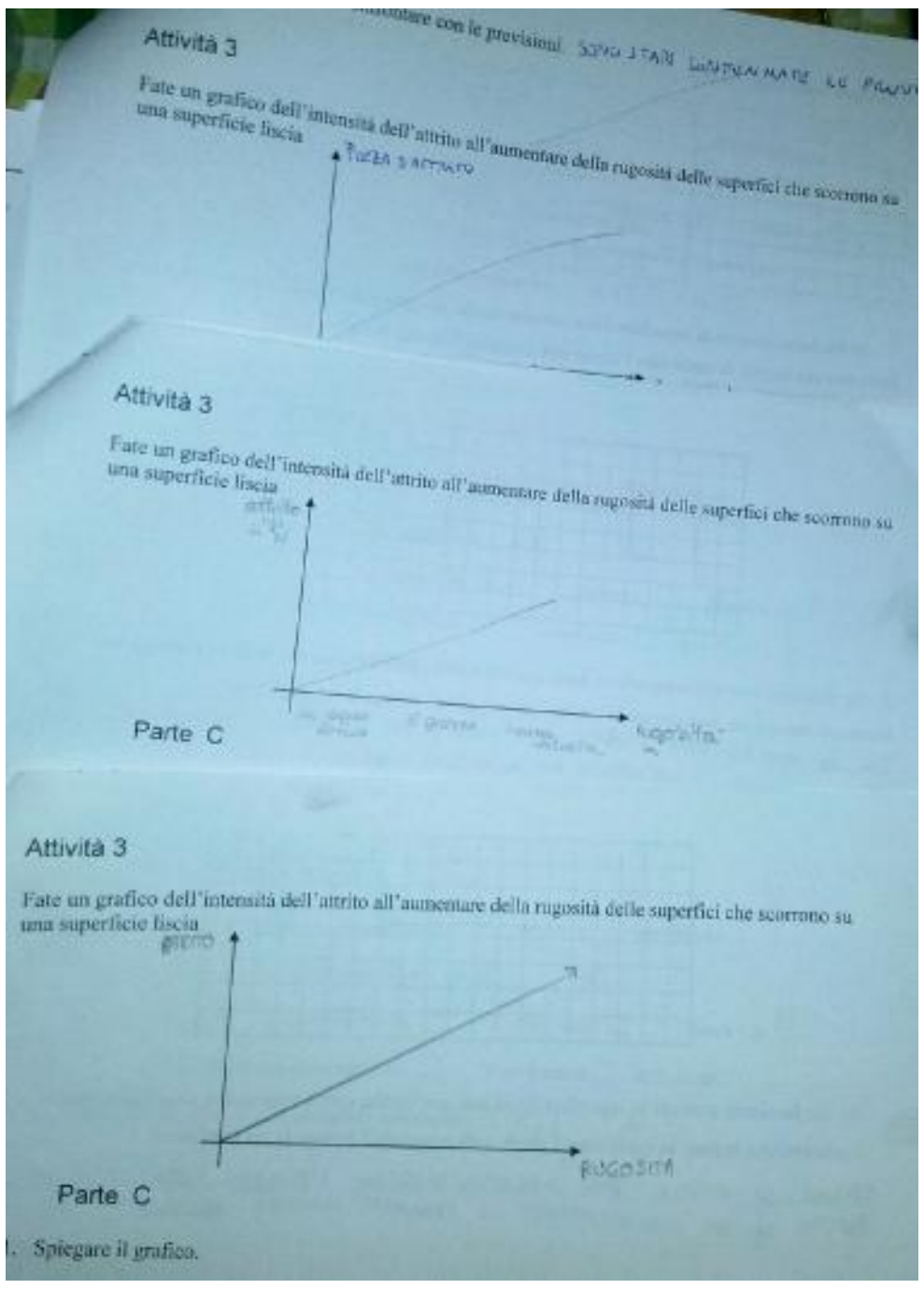

Figure 2. All students draw essentially a direct proportional behaviour for friction force intensity vs. surface roughness

Almost all students sketched a graph of friction force intensity vs. surface roughness very similar to those found in an analogous study in higher education [2].

Some remarks can be done by observing the graphs showed in fig. 2. All students seemed to identify an increasing dependence with a linear function. Moreover, recent observations with wooden surfaces and sandpapers were extended to region of roughness not observed, without any doubt that could exist a different behaviour (specific questions on this point were inserted in the worksheet and the answers were very clear).

Afterwards, metal blocks (showed in fig. 3) with rough and smooth surfaces were examined and a new prevision requested. All students were consistent with the general dependence drawn in their previous graphs, but this time the following test failed in confirming their previsions. Thus, students were engaged in a cognitive conflict.

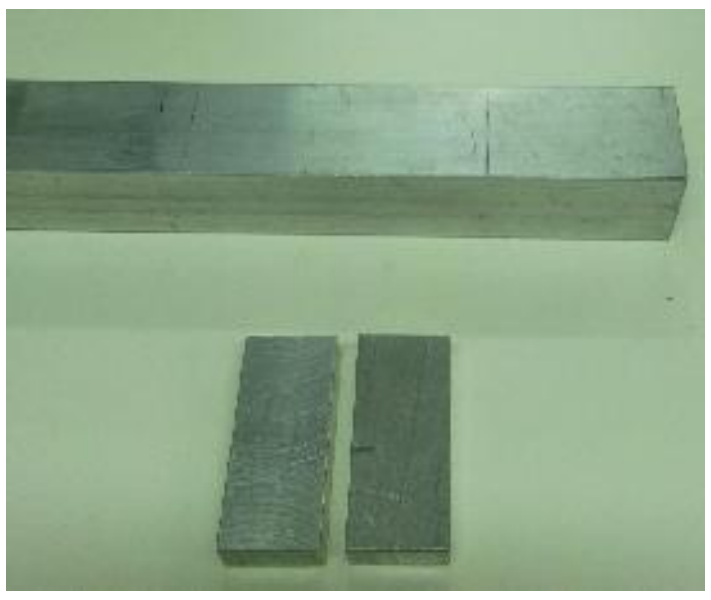

Figure 3. Metal blocks with different roughness were examined.

A hint for resolving this conflict was given by the final activity. Students started from a macroscopic perspective in which friction is simply due to mechanical interactions of the atoms (interlocking or intertwining, rubbing or hitting bumps and valleys). A sheet of paper and a transparency were made sliding in different conditions (see fig. 4) suggesting that the correct interaction involved in friction is the electromagnetic one. Thus, students were introduced to the microscopic underpinnings of friction.

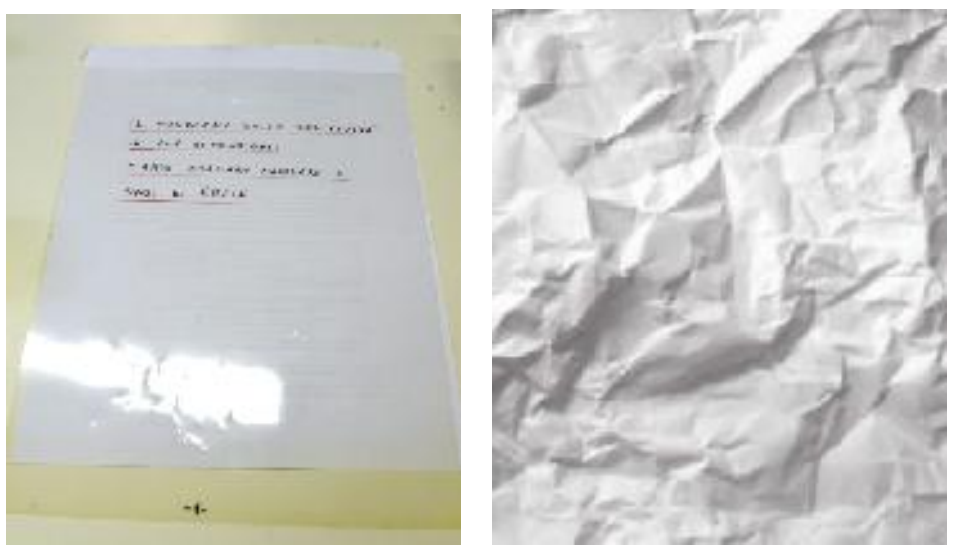

Figure 4. A transparency slides with difficulty on a sheet of paper electrostatically charged by rubbing (on the left), but it slides easier if the sheet was crumpled and flattened with hands (on the right).

Moreover, the relation between the friction force intensity and the area of contact spontaneously emerged as well as the possibility that the apparent area of contact can be different from the real one.

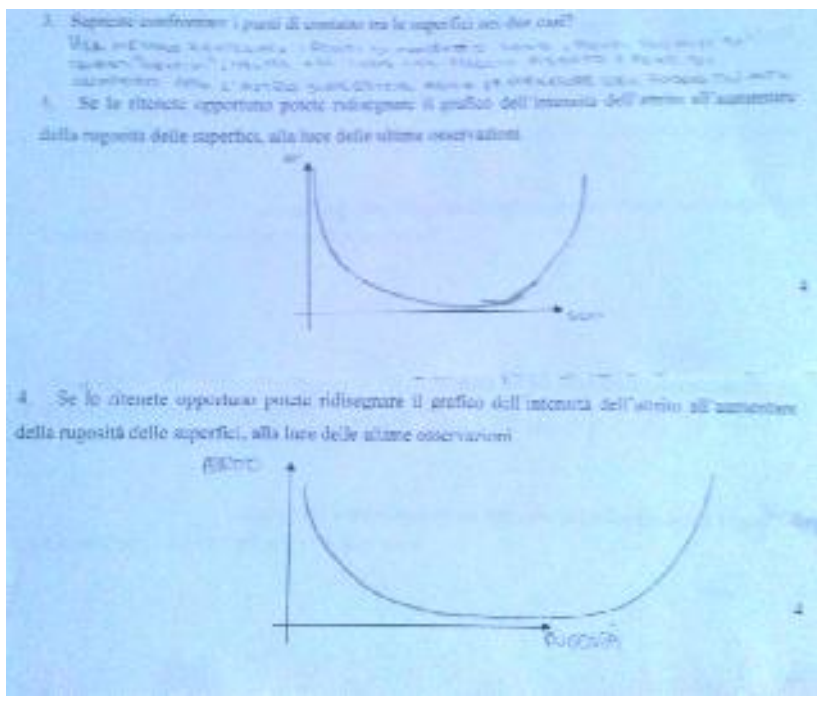

Figure 5. Revised graph of friction force versus roughness of the sliding surfaces according to observations.

Finally, students could make a new graph of friction force vs. roughness, if they believed it was necessary. The most part of students (74%) revised their graph, but only few of them (4%) were able to sketch the one consistent with all previous observations (fig. 5) and according with the evidence from phenomenology [7].

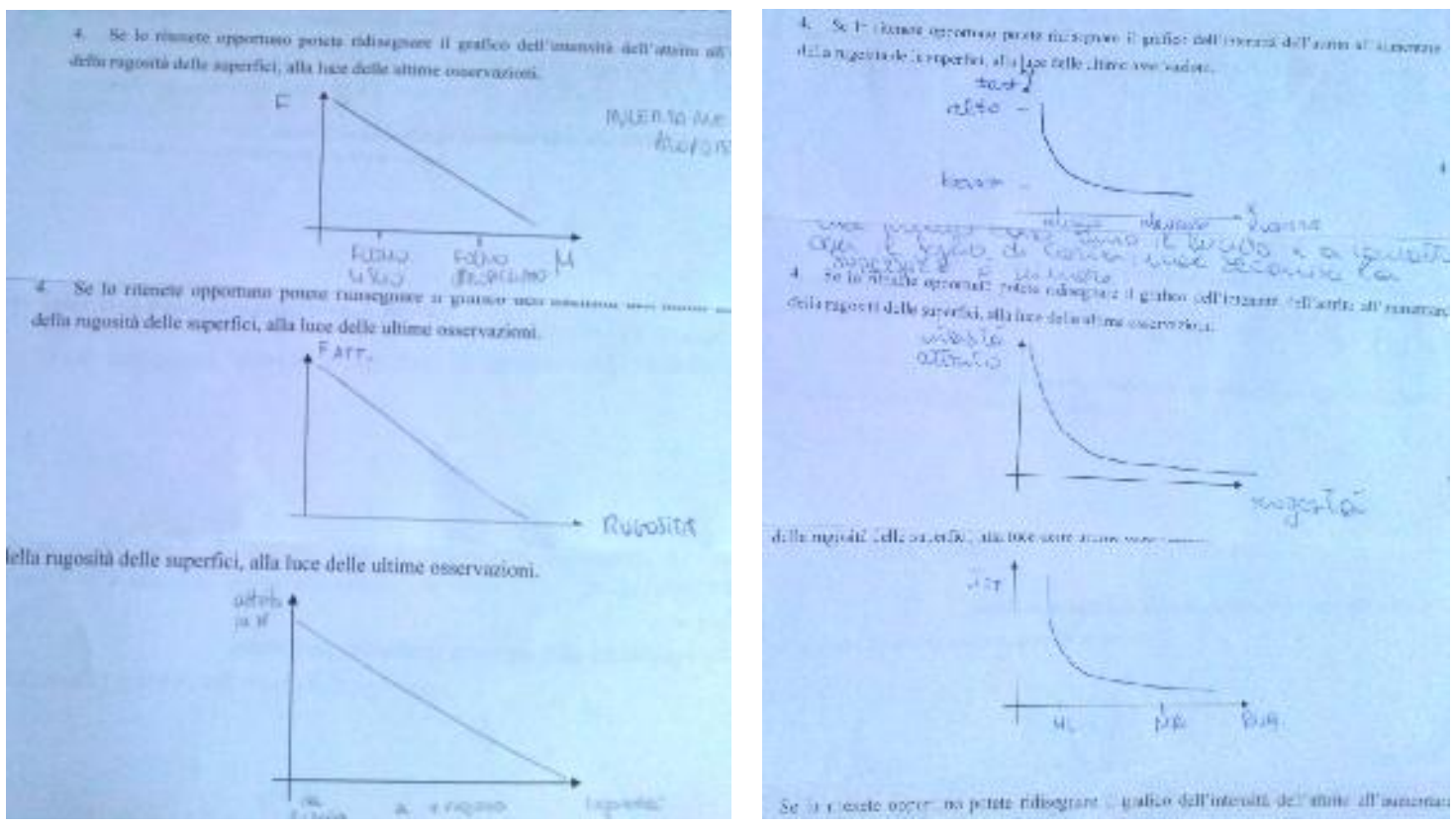

Figure 6. Other graphs can be equally divided in linear decreasing functions (on the left) and a decreasing hyperbolic function (on the right)

The most part of students gave more relevance to the last experience in laboratory and avoid to connect last and previous observations in a consistent graph. As shown in fig. 6, graphs show only a decreasing behaviour for friction force intensity.

## A quantitative experience on friction

Almost all qualitative observations were tested in an experiment. The focus was on static friction and its dependence from load, sliding surface area and roughness. Students worked in small groups (3 or 4 components). Each group tested a different aspect encountered in the qualitative exploration. In particular, static friction dependence from sliding surfaces of the same material, from apparent contact area and from load were investigated both for wood and stainless steel. Each student had to produce a report on his or her experiment in which coefficients of static friction were determined.

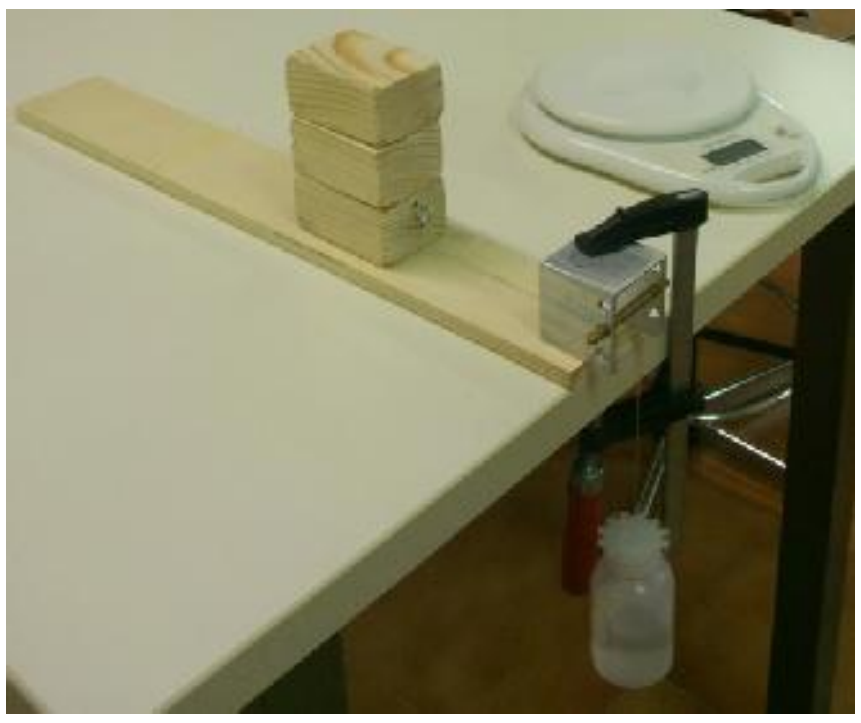

Figure 7. Wooden blocks could slide on a wooden plane. The experiment could be repeated with a different load, apparent area of contact or surface roughness.

Experiments consisted mainly of blocks (woody or metallic) that could slide on a plane (woody or metallic) pulled by a force supplied by the weight of an hanging bottle containing water. Water was added up to determine the least amount that set in motion the body (see fig. 7). The intensity of the force was computed by using the measurement of the mass of the bottle made with a balance.

In figure 8, the experimental set-up is showed for stainless steel blocks. In this case, it was possible to vary the position of the support wire that pulled the block so that the applied force was always parallel to the sliding plane in such a way that the force pulling the blocks had the same intensity as the weight of the hanging bottle.

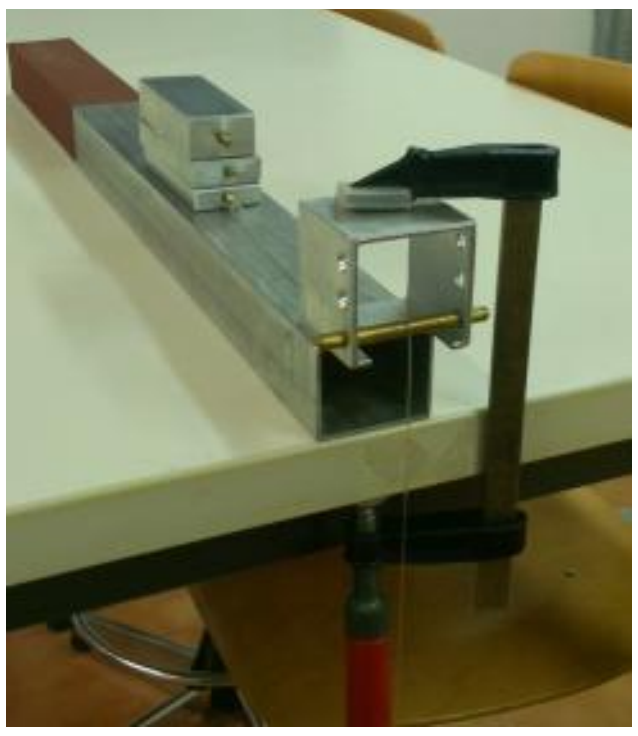

Figure 8. Experimental set-up for metal blocks.

An unusual aspect of this kind of experiments was the necessity of identifying the intensity of the force and the relative uncertainty by using two "limit" measurements (maximum force without movement, minimum one with movement), like shown in fig. 9.

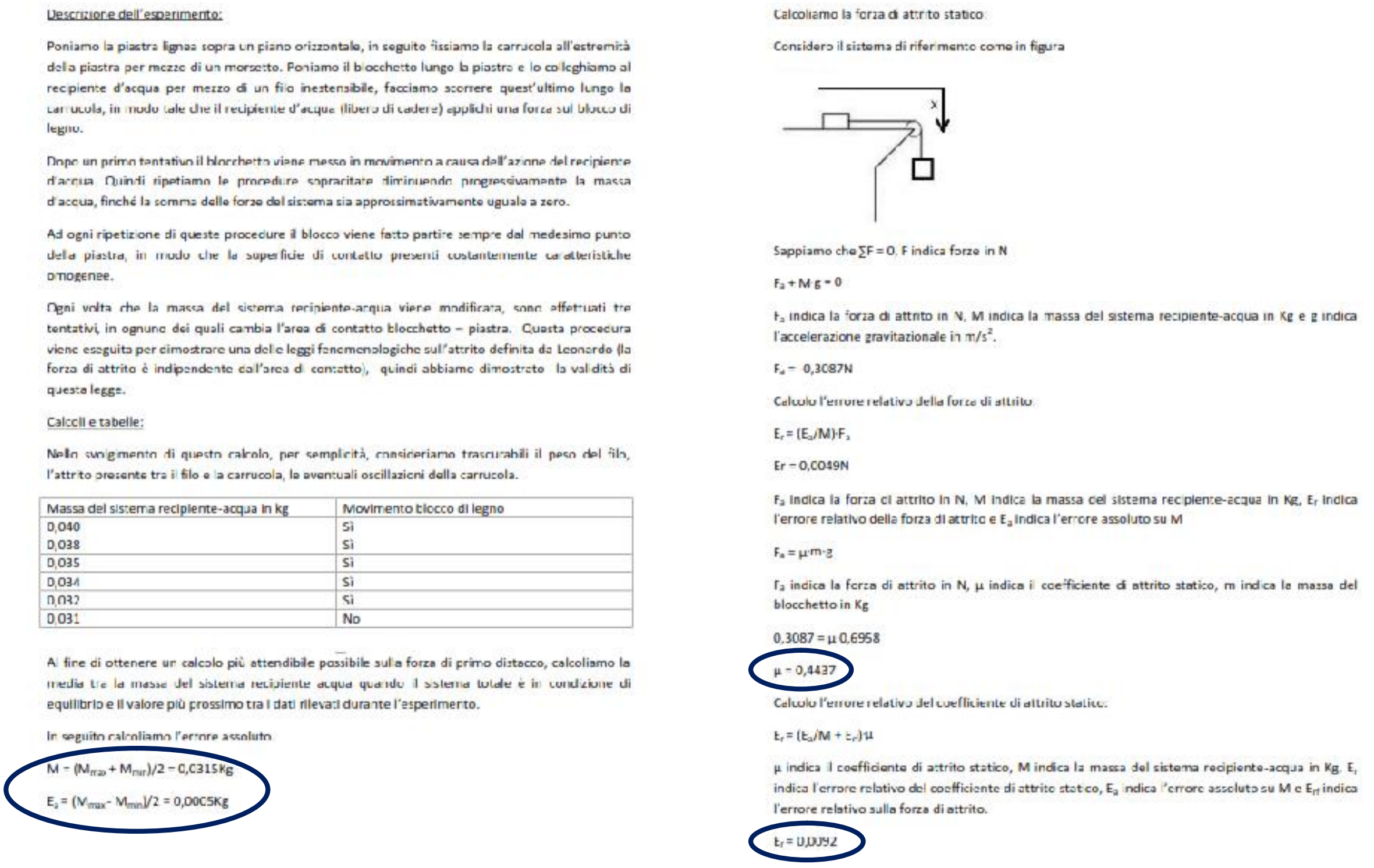

Descrizione dell'esperimento:

Poniamo la piastra lignea sopra un piano orizzontale, in seguito fissiamo la carrucola all'estremità della piastra per mezzo di un morsetto. Poniamo il blocchetto lungo la piastra e lo colleghiamo al recipiente d'acqua per mezzo di un filo inestensibile, facciamo scorrere quest'ultimo lungo la carrucola, in modo tale che il recipiente d'acqua (libero di cadere) applichi una forza sul blocco di legno.

Dopo un primo tentativo il blocchetto viene messo in movimento a causa dell'azione del recipiente d'acqua. Quindi ripetiamo le procedure sopracitate diminuendo progressivamente la massa d'acqua, finché la somma delle forze del sistema sia approssimativamente uguale a zero.

Ad ogni ripetizione di queste procedure il blocco viene fatto partire sempre dal medesimo punto della piastra, in modo che la superficie di contatto presenti costantemente caratteristiche omogenee.

Ogni volta che la massa del sistema recipiente-acqua viene modificata, sono effettuati tre tentativi, in ognuno dei quali cambia l'area di contatto blocchetto – piastra. Questa procedura viene eseguita per dimostrare una delle leggi fenomenologiche sull'attrito definita da Leonardo (la forza di attrito è indipendente dall'area di contatto), quindi abbiamo dimostrato la validità di questa legge.

Calcoli e tabelle:

Nello svolgimento di questo calcolo, per semplicità, consideriamo trascurabili il peso del filo, l'attrito presente tra il filo e la carrucola, le eventuali oscillazioni della carrucola.

| Massa del sistema recipiente-acqua in kg | Movimento blocco di legno |
|---|---|
| 0,040 | Sì |
| 0,038 | Sì |
| 0,035 | Sì |
| 0,034 | Sì |
| 0,032 | Sì |
| 0,031 | No |

Al fine di ottenere un calcolo più attendibile possibile sulla forza di primo distacco, calcoliamo la media tra la masse del sistema recipiente acqua quando il sistema totale è in condizione di equilibrio e il valore più prossimo tra i dati rilevati durante l'esperimento.

In seguito calcoliamo l'errore assoluto

$M = (M_{max} + M_{min})/2 = 0{,}0315 Kg$

$E_a = (M_{max} - M_{min})/2 = 0{,}0005 Kg$

Calcoliamo la forza di attrito statico

Considero il sistema di riferimento come in figura

Sappiamo che $\sum F = 0$, F indica forze in N

$F_a + M \cdot g = 0$

$F_a$ indica la forza di attrito in N, M indica la massa del sistema recipiente-acqua in Kg e g indica l'accelerazione gravitazionale in $m/s^2$.

$F_a = 0{,}3087 N$

Calcolo l'errore relativo della forza di attrito

$E_r = (E_a/M) \cdot F_a$

$Er = 0{,}0049 N$

$F_a$ indica la forza di attrito in N, M indica la massa del sistema recipiente-acqua in Kg, $E_r$ indica l'errore relativo della forza di attrito e $E_a$ indica l'errore assoluto su M

$F_a = \mu \cdot m \cdot g$

$F_a$ indica la forza di attrito in N, μ indica il coefficiente di attrito statico, m indica la massa del blocchetto in Kg

$0{,}3087 = \mu \, 0{,}6958$

$\mu = 0{,}4437$

Calcolo l'errore relativo del coefficiente di attrito statico:

$E_r = (E_a/M + E_{rf})\mu$

μ indica il coefficiente di attrito statico, M indica la massa del sistema recipiente-acqua in Kg, $E_r$ indica l'errore relativo del coefficiente di attrito statico, $E_a$ indica l'errore assoluto su M e $E_{rf}$ indica l'errore relativo sulla forza di attrito.

$E_r = 0{,}0092$

Figure 9. The determination of static friction coefficient in a student's report.

Leonardo's laws (independence of friction from sliding surface area and linear dependence from load) were verified for steel. In the case of wooden blocks, measurements were not always in accord with Leonardo's laws and sometimes gave different results when repeated. Teachers and students agreed that the problem was in the light load and in the lack of hardness in the material of the surface. During the experiment, wooden surface roughness could show an increasing wear.

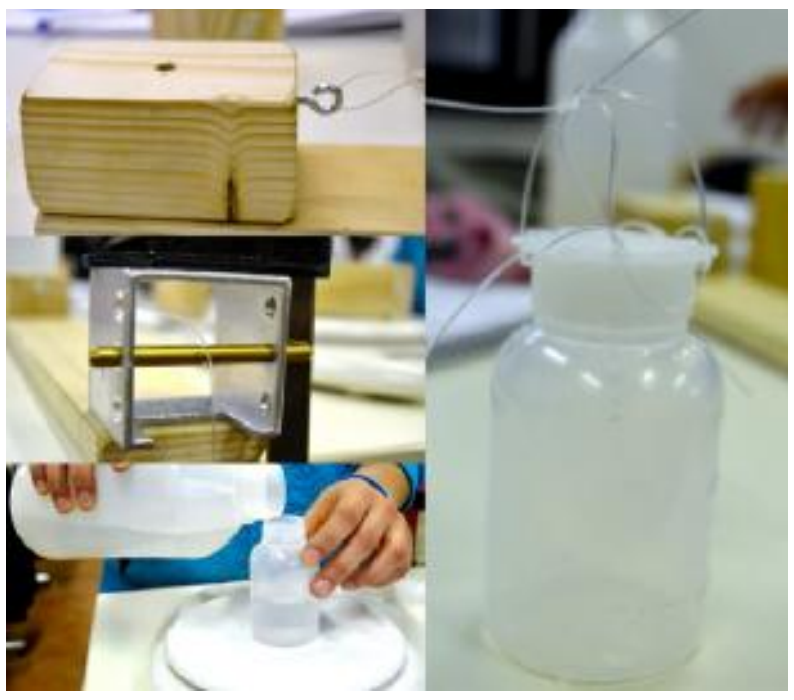

Figure 10. A collage of images presented in a report in which the student resumed the main aspects of her laboratory on friction.

Despite of the general interest and involvement showed by students (see fig. 10), many discussions in the groups and teachers' clarification, which seemed effective at the moment, some severe and unexpected learning difficulties arose in the reports presented by students some weeks later. In particular, some students were unable to evaluate the friction force intensity and a larger number gave no correct force's uncertainty (results are summarized in fig. 11).

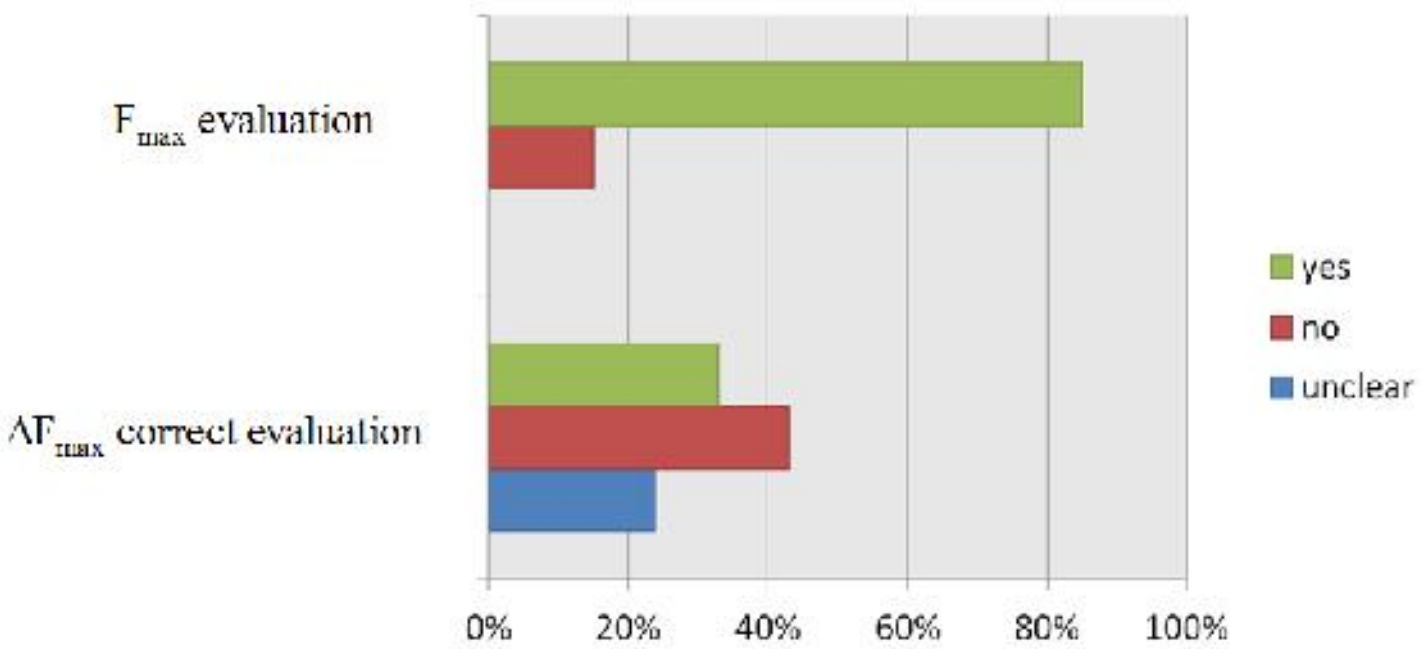

Figure 11. Percentage of correct evaluations of the maximum intensity of static friction and its uncertainty. The most common error is to evaluate uncertainty by considering only the sensibility of the measuring instrument, the balance in this case. The unclear cases consist in evaluations without explanations or explicit computations.

Moreover, a couple of students obtained inconsistent measurements respect to qualitative observations or well-established laws on friction. They seemed unaware and were not able to recognize that a problem exists.

## Conclusions

All students were very involved in the inquiry-based introductory activity and seemed to realize with care the successive measurements. This approach had improved students' understanding of friction, by discovering a rich phenomenology closely related to the structure of matter and its interactions at the nanoscale.

However, a preliminary analysis of their reports shows some learning difficulties, based mainly on the lacking or weakness about previous basic concepts in physics laboratory or mathematics which still do not seem well assimilated:

- students had real difficulties to imagine a functional dependence different from

direct and inverse proportion**.** Furthermore, many students confused the contrary behaviour of the direct proportion with the decreasing linear dependence, as shown in fig. 6.

- few students were able to merge different observed behaviours in a unique phenomenological graph.
- despite of their long experience in physics laboratory with their teacher (more than 20 experiments realized in small groups in two school years) and all explanations and clarifications obtained in laboratory, many students were unable to evaluate an uncertainty different from instrumental one in an unusual measure.

These difficulties suggest that in practice some skills developed in the physics laboratory (but also fundamental topics in mathematics) are not used properly in an inquiry approach, i.e. they are not acquired correctly and completely. More research on this topic is necessary.

A further step in the quantitative exploration of friction can be made by realizing an experience on kinetic friction. The same experimental set-up can be used in the dynamic case (hanging more mass), if the motion of the block is recorded by a camera. The coefficient of kinetic friction can be measured by elaborating images through a video analysis tool (e. g. an open source software like Tracker [8]) and Coulomb's law for friction can be verified. Also in this case, it is essential that students have a more solid training in lab as prerequisite.